\documentclass[aps,prl,reprint,superscriptaddress,longbibliography,floatfix]{revtex4-1}

\usepackage{amsmath,amssymb} 
\usepackage{graphicx}
\usepackage{verbatim}

\begin{document}

\title{Dielectric Modulation of Ion Transport near Interfaces}

\author{Hanne S. Antila}
\affiliation{Department of Materials Science \& Engineering, Northwestern
  University, Evanston, IL 60208, USA}

\author{Erik Luijten}
\affiliation{Departments of Materials Science \& Engineering, Engineering
  Sciences \& Applied Mathematics, and Physics \& Astronomy, Northwestern
  University, Evanston, IL 60208, USA}
\email[Corresponding author: ]{luijten@northwestern.edu}

\date{November 29, 2017}

\begin{abstract}
  Ion mobility and ionic conductance in nanodevices are known to deviate from
  bulk behavior, a phenomenon often attributed to surface effects. We
  demonstrate that dielectric mismatch between the electrolyte and the surface
  can qualitatively alter ionic transport in a counterintuitive
  manner. Instead of following the polarization-induced modulation of the
  concentration profile, mobility is enhanced or reduced by changes in the
  ionic atmosphere near the interface and affected by a polarization force
  parallel to the surface. In addition to revealing this mechanism, we explore
  the effect of salt concentration and electrostatic coupling.
\end{abstract}


\maketitle

Understanding ion mobility and ionic conductance is of fundamental importance
in fields ranging from biology to energy conversion, describing phenomena as
diverse as ion channels~\cite{gouaux05} and fuel cells~\cite{kreuer14}. The
foundation for this understanding was laid more than a century ago by
Kohlrausch~\cite{kohlrausch1879,kohlrausch07,atkins2006}, who observed that
the molar conductivity $\Lambda_{m}$ of electrolytes decreases with increasing
salt concentration~$c$, $\Lambda_{m}=\Lambda_{0}-A\sqrt{c}$.  Debye and
H{\"u}ckel~\cite{debye23b,debye27}, and Onsager~\cite{onsager27,onsager31}
connected this concentration dependence to the counterion atmosphere
surrounding moving ions. This atmosphere, which has a size related to the
concentration via the Debye length~$\lambda_{\rm D} \propto 1/\sqrt{c}$,
exerts two types of forces on the central ion, the electrophoretic force and
the relaxation force. The electrophoretic force arises from the modification
of the viscous drag on the central ion by solvent molecules that are pulled in
the opposite direction by the counterions. The relaxation force is a
consequence of the asymmetry of the ionic atmosphere under a driving
field. The atmosphere around a moving ion is continuously being rebuilt---a
process that takes finite time and causes the center of mass of the atmosphere
to lag behind the central ion.  Due to this asymmetry, the ion cloud exerts a
Coulombic force on the central ion, slowing down its motion.

The original derivation of Debye, H{\"u}ckel, and Onsager relies on various
simplifications. Subsequent conductance theories~\cite{fuoss57,fuoss78} more
accurately take into account non-idealities, such as ion association, as well
as the coupling between relaxation and electrophoretic effects, notably the
modification of the latter by the asymmetry of the ionic atmosphere. These
corrections result in higher-order terms in the concentration and extend the
validity of the theory to a wider concentration range. Nevertheless, the
effect of concentration on molar conductivity remains qualitatively unchanged,
namely that ion mobility decreases as concentration increases.

Under nanoscale confinement, ion mobilities~\cite{duan10} and
conductances~\cite{stein04} are known to deviate from bulk-like
behavior. Moreover, such devices also exhibit other special transport
properties, e.g., ion selectivity~\cite{nishizawa95,cervera06} and
rectification~\cite{cervera06}. These deviations from bulk behavior are often
attributed to surface effects and to the high surface-to-volume ratio
characteristic of nanodevices. For example, net positive surface charge will
attract excess negative ions into a nanopore. At low concentrations, this will
enhance the conductance compared to the bulk~\cite{stein04}. Conversely,
surface charge has been predicted to increase water viscosity and thereby
decrease ion mobility near the surface~\cite{qiao05b}, where the specific ion
mobility depends on the sign of the surface charge~\cite{qiao05a} and on ionic
characteristics.

Yet another effect concerns the permittivity. Materials used in synthetic
nanoscale devices range from dielectric to metallic, so that the surface
polarization induced at the fluid--solid interface may influence ion
transport. In addition to dielectric exclusion~\cite{buyukdagli11} of ions
from nanopores, the dielectric properties of a pore have been predicted to
enhance ion selectivity~\cite{boda06}. Intriguingly, recent
calculations~\cite{zhang11} have raised the possibility that the permittivity
of the pore surface can be used to tune ionic rectification in conical
nanopores. Thus, along with surface charge and pore size, permittivity
potentially provides an additional parameter for achieving a high degree of
control over ion movement. Nevertheless, to the best of our knowledge,
hitherto all studies of the effect of surface polarization on ionic
conductivity have concentrated on the distribution, number, or type of ions in
the pore~\cite{mamonov03,zhang11,tagliazucchi13,balme15}, whereas their
mobility, an essential factor in the overall ionic conductivity, has been
assumed to be independent of the dielectric properties of the nanodevice
surface.

Here, we address this knowledge gap and demonstrate that the mobility of ions
near a surface indeed can be controlled by tuning the dielectric mismatch
between the wall and the solvent. We relate the origins of this effect to
modifications the surface polarization induces in the counterion
atmosphere, and in the related relaxation force.  Molecular dynamics (MD)
simulations permit a microscopic view of ion mobility and counterion clouds as
a function of ion distance to the interface. To include fluctuation and
correlation effects all ions are treated explicitly, whereas both the solvent
and the surface are modeled as dielectric continua.  The use of a
coarse-grained model allows us to incorporate dielectric effects into the
simulations, and to track the movement of ions for long enough times ($10^{9}$
simulation steps, corresponding to more than 5~ms) to allow reliable
extraction of the mobility and corresponding forces.

\begin{figure}
\includegraphics[scale=0.41]{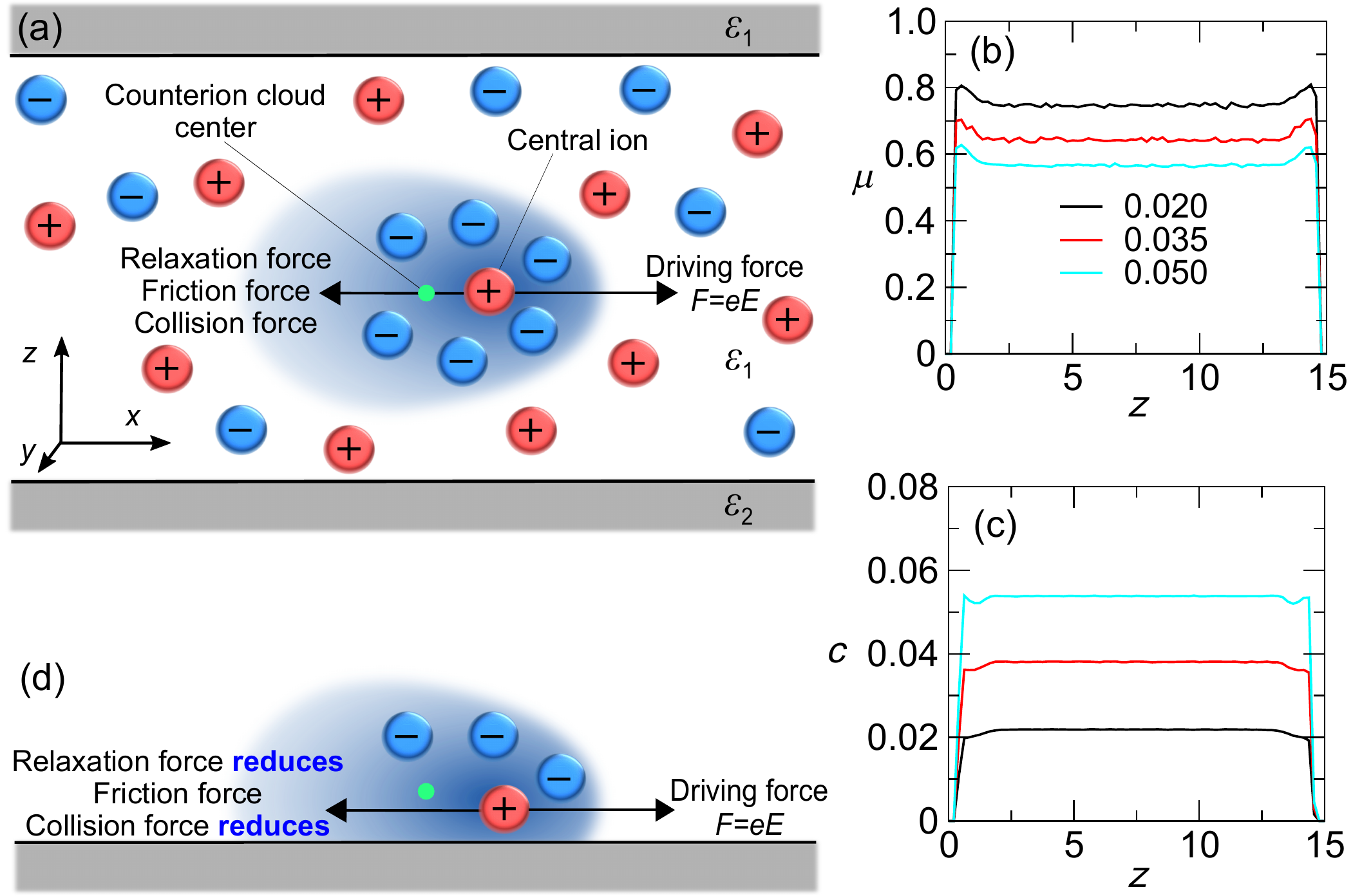}
\caption{Schematic of the simulation system and the forces exerted on an ion
  accompanied by the mobilities [unit $\sigma^{2}e/(k_{\rm B}T\tau)$] and ion
  concentrations [unit $\sigma^{-3}$] without dielectric mismatch, i.e.,
  $\varepsilon_{1}=\varepsilon_{2}$. a)~Simulation set-up. Ions are confined
  between two plates and an electric field is applied parallel to the
  surface. Relaxation force, collision force, and friction force oppose the
  ion motion. b)~Mobility~$\mu$ of ions in the absence of dielectric mismatch
  ($\Delta = 0$) as a function of distance~$z$ to the lower surface. Colors
  denote different ion concentrations~$c$. c)~Ion concentration profiles for
  the same systems as in panel~c. d)~Forces on an ion residing near an
  interface without dielectric mismatch. The surface distorts the counterion
  atmosphere, thereby reducing the forces that slow down the ion.}
\label{fig:schematic}
\end{figure} 

We adopt the restricted primitive model~\cite{luijten02a}, modeling ions as
monovalent ($q=\pm e$), purely repulsive shifted-truncated Lennard-Jones
spheres of mass $m$ and diameter $\sigma$, which we choose as our unit of
length. For hydrated ions, $\sigma$ is approximately 0.7~nm. We employ a
parallel-plate geometry of width and length $L_x = L_y = 15\sigma$,
periodically replicated in both dimensions. The top and bottom surfaces are
separated by $L_z = 15\sigma$. The upper surface has the same dielectric
constant as the solvent, $\varepsilon_1$, whereas the lower surface has
dielectric permittivity $\varepsilon_2$. This geometry makes it possible to
account for the effects of the complex surface polarization patterns via image
charges~\cite{neumann1883}.  To accommodate the image charges, the height of
the actual simulation cell is doubled, and all electrostatic interactions are
computed via 3D PPPM with accuracy $10^{-5}$, and a slab correction
accompanied by $60\sigma$-thick vacuum layer. We use the dielectric mismatch
$\Delta = (\varepsilon_{1}-\varepsilon_{2})/(\varepsilon_{1}+\varepsilon_{2})$
to describe the magnitude and sign of the image charge: $\Delta = 1$ for a
low-permittivity surface that results in repulsive surface polarization,
$\Delta = 0$ for an interface with no dielectric mismatch, and $\Delta = -1$
for a high-permittivity surface with attractive surface polarization.  We use
a timestep of 0.01$\tau$, where $\tau=\sqrt{m\sigma^{2}/\varepsilon_{LJ}}$,
$\varepsilon_{\rm LJ}=k_{\rm B}T/1.2$ is the Lennard-Jones coupling constant,
$T$ denotes the absolute temperature, and $k_{\rm B}$ is Boltzmann's
constant. Following the convention in polyelectrolyte
simulations~\cite{hsiao06}, we employ an enhanced Bjerrum length
$l_{\rm B}=3\sigma$. Whereas this enhances the electrostatic effects, we will
demonstrate that our findings hold at lower coupling strength as well. Unless
stated otherwise, the ion concentration is $c = 0.02\sigma^{-3}$
(corresponding to $0.1$M).

The simulation setup and the forces affecting the movement of ions are
depicted in Fig.~\ref{fig:schematic}a.  Ions are driven by an external field
$E=0.4k_{\rm B}T/(e\sigma)$ in the $x$-direction.  This field strength lies
within the linear response regime, and is counteracted by the relaxation
force, frictional forces, and the collision force. The friction force (viscous
drag) exerted by the solvent on individual ions is captured by a Langevin
thermostat, applied in the system with damping constant
$\gamma=m\tau^{-1}$~\cite{note-units}. The short-range drag arising from
interacting hydration shells of ions that pass each other is represented by
Lennard-Jones collisions between ions.  Since our simulations do not
incorporate hydrodynamics, the ions do not experience a long-range
electrophoretic force. However, as this force has the same functional
dependence on salt concentration as the explicitly included relaxation
force~\cite{onsager31}, this does not qualitatively affect Kohlrausch's
law. Moreover, as we will discuss below, our findings regarding the role of
surface permittivity are equally unaffected.  The ion mobility
[$\mu=\langle v\rangle/(Eq)$] is determined by the balance of these force
components, and obtained by averaging the instantaneous velocity
$\langle v \rangle$ of ions.

To establish a reference system, we first explore ion mobilities
(Fig.~\ref{fig:schematic}b) and the underlying ion concentration profiles
(Fig.~\ref{fig:schematic}c) in a channel without dielectric mismatch. As
expected, the ion mobility decreases as concentration increases, in
qualitative agreement with Kohlrausch's law.  However, the profiles are not
uniform, displaying an increase in the ion mobilities near the surfaces for
all concentrations (we examined $c \leq 0.1\sigma^{-3}$).  Indeed, this
mobility increase reflects the important role of the counterion atmosphere in
ion conductivity. The presence of the wall perturbs the ion cloud and leads to
a decrease both in the electrostatic relaxation force and in ion--ion
collisions (Fig.~\ref{fig:schematic}d), which in turn increases the mobility
near the interface. Even though this can readily be produced in a simple MD
simulation, we are unaware of prior previous reports on this effect.

\begin{figure}
\includegraphics[scale=0.67]{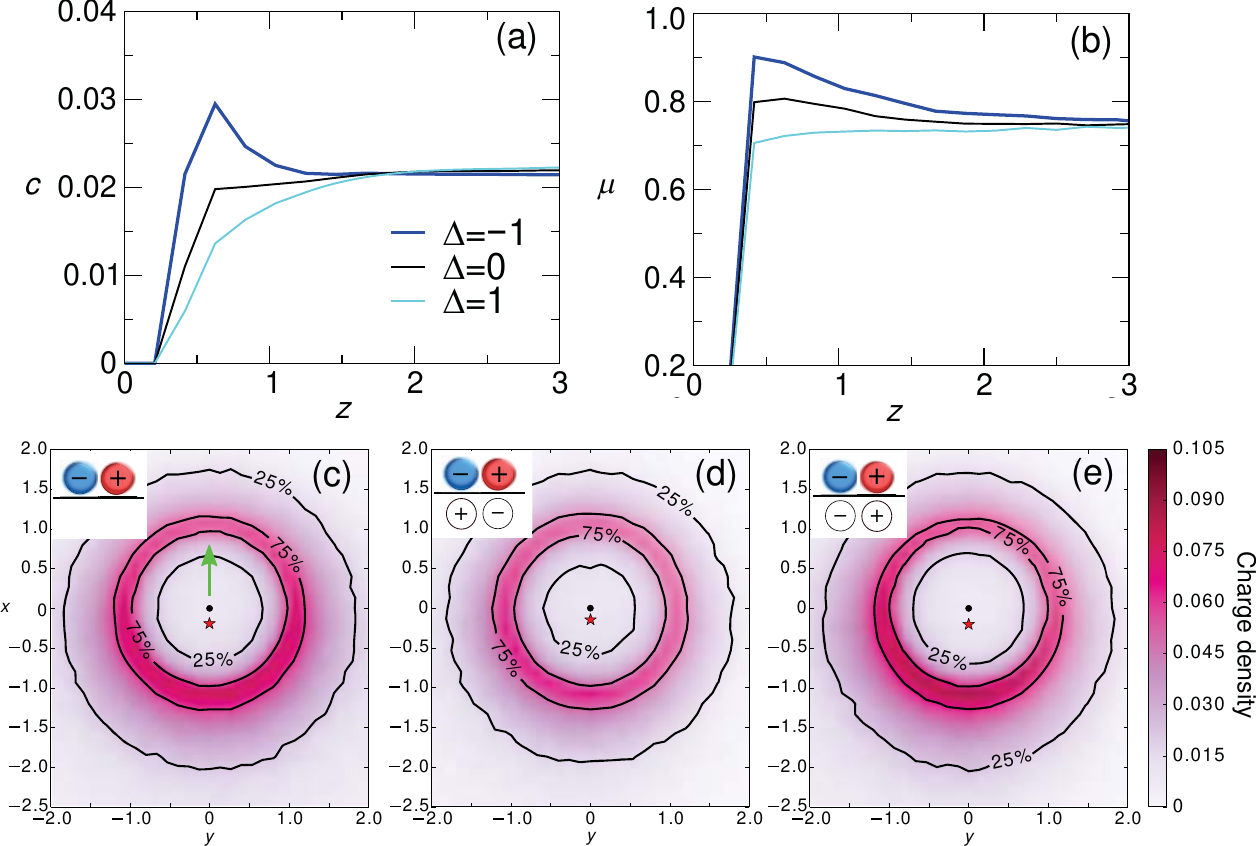}
\caption{Effect of dielectric mismatch $\Delta$ on ion distributions and
  mobilities, at a bulk concentration $c = 0.020\sigma^{-3}$.  (a)~Ion
  concentrations as function of distance to the bottom wall for different
  values of $\Delta$. (b)~Corresponding mobility of ions (colors as in
  panel~a). (c--e)~2D charge densities around negative ion within cutoff of
  2$\sigma$ from the surface: c)~$\Delta=0$, d)~$\Delta=-1$,
  e)~$\Delta=1$. Small black circle marks the central ion position, green
  arrow shows the direction of movement, and a red star labels the center of
  the charge distribution. Contours of 25\% and 75\% of maximum charge density
  are shown to demonstrate shape of ion atmosphere. Charge of the central ion
  is not taken into account in the visualization; the
  depletion of charge around the central ion is caused by the
  $z$-cutoff. Insets are schematics clarifying the image charge effect on
  ion--ion interactions. Units as in Fig.~\ref{fig:schematic}.}
\label{fig:ionatmosphere}
\end{figure}

The situation becomes more complex when surface polarization is taken into
account. Figure~\ref{fig:ionatmosphere}a shows the expected build-up of ions
near an attractive, high-permittivity surface ($\Delta = -1$) and depletion
near a low-permittivity material ($\Delta = 1$,
Fig.~\ref{fig:ionatmosphere}a). Based on Kohlrausch's law, and our
observations in Fig.~\ref{fig:schematic}b,c, the mobilities should
consequently decrease near a surface with $\Delta = -1$ and increase near a
surface with $\Delta = 1$. Surprisingly, we observe the
opposite. Figure~\ref{fig:ionatmosphere}b shows that near a high-permittivity
surface the interfacial mobility is enhanced compared to a system without
dielectric mismatch ($\Delta = 0$), whereas a surface with low dielectric
constant decreases the mobility.

We hypothesize that this remarkable behavior results from changes in the ionic
atmosphere. Indeed, in bulk electrolytes such changes are known to affect ion
mobilities.  For example, in the Wien effect~\cite{wien28,onsager57,luijten13}
electrolyte mobility increases in high fields because the fast movement of the
ions prevents the formation of the counterion cloud.  Similarly, the
Debye--Falkenhagen effect~\cite{debyefalk28,falkenhagen1934} describes how in
high-frequency AC fields the fast, continuous switching of the direction of
the ion movement suppresses the asymmetry of the ionic atmosphere, so that the
relaxation force vanishes.

Accordingly, we examine the effect of surface polarization on counterion
atmospheres surrounding ions in the interfacial region.
Figure~\ref{fig:ionatmosphere}c depicts the shape and net charge density of
the ionic cloud in the absence of surface polarization. It confirms the
distortion of the cloud in the direction of motion, with its center of mass
located \emph{behind} the central ion. Attractive polarization ($\Delta = -1$,
Fig.~\ref{fig:ionatmosphere}) weakens the overall counterion cloud and
simultaneously suppresses its asymmetry. This in turn diminishes the
relaxation force, resulting in the speed-up observed in
Fig.~\ref{fig:ionatmosphere}b. The inset illustrates the underlying mechanism,
which is phrased most concisely in term of the image charges that represent
the induced surface polarization patterns. Counterions in the cloud are
repelled by the image of the central ion.  This weakens the ion--ion
interactions and thereby not only diminishes the net charge of the ionic
cloud, but also makes it more symmetric, since the range of the ionic
atmosphere is connected to the relaxation time needed to rebuild it.

Conversely, repulsive surface polarization ($\Delta = 1$,
Fig.~\ref{fig:ionatmosphere}e) enhances both the intensity and asymmetry of the
ionic atmosphere. This leads to an increase in the relaxation force, and to a
slow-down of ions close to the interface, supporting the mobility
profile observed in Fig.~\ref{fig:ionatmosphere}b. The interaction between
ions and their own images is now repulsive, whereas the secondary interaction
between an ion and the image of its countercharge is attractive. This leads to
enhanced ion--ion attraction and to the elevated net charge density around
an ion residing near a low-dielectric surface.

The modulation of ion--ion interactions by polarizable
surfaces~\cite{nadler03} and the consequent changes in ionic
atmosphere~\cite{buyukdagli11,zwanikken13} near interfaces have been predicted
before. Experimental support for the weakening of ion--ion interactions near a
high-permittivity material is provided by the observation of enhanced
dissociation of a weak electrolyte, leading to more free charge carriers and
an increase in conductivity~\cite{korobeynikov05}. However, to the best of our
knowledge, the modulation of ion mobilities by polarizable interfaces through
changes in the ionic atmosphere has not been reported before.

\begin{figure}
  \includegraphics[scale=0.6]{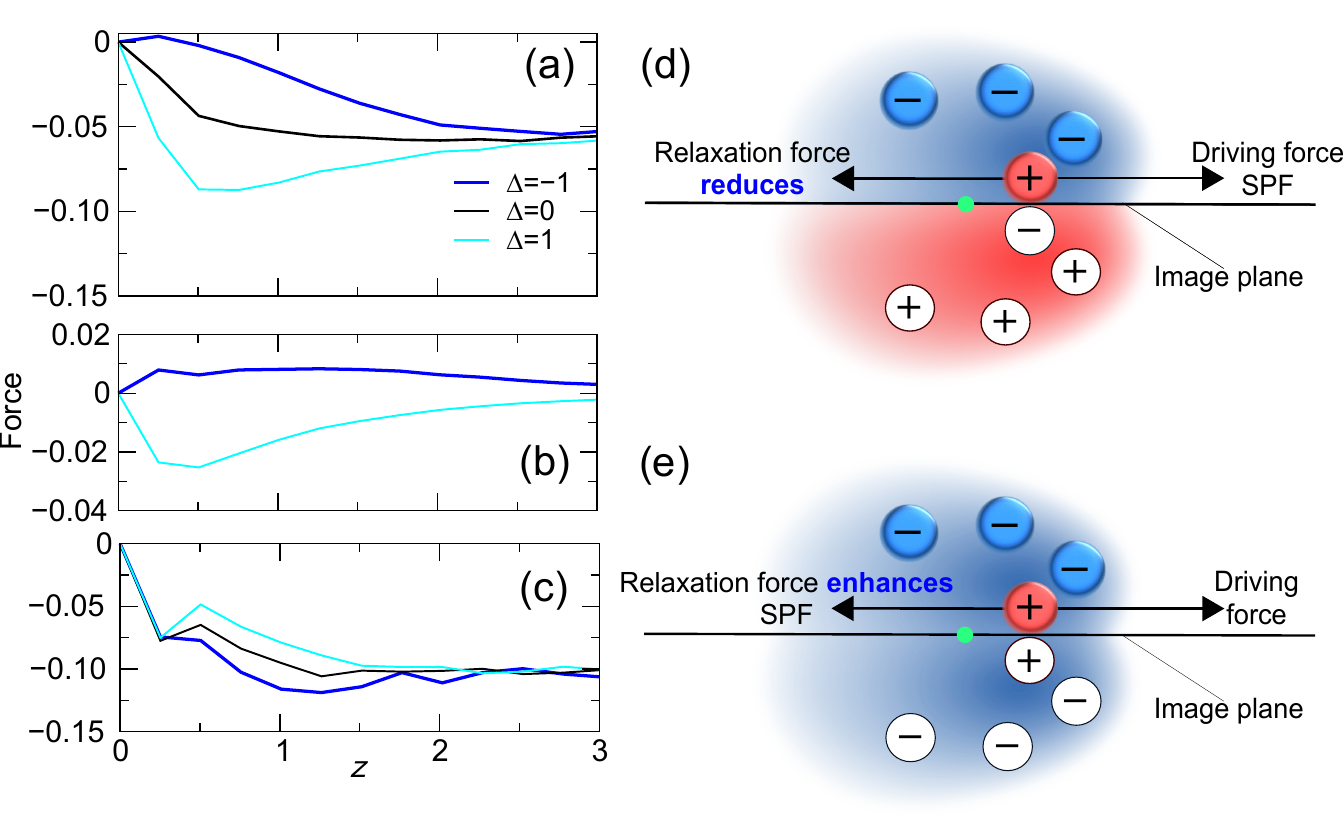}
  \caption{Forces (in the direction of motion; unit $k_{\rm B}T/\sigma$)
    exerted on the ions for the systems of
    Fig.~\ref{fig:ionatmosphere}. (a)~Total relaxation force as a function of
    distance to the channel wall. (b)~Surface polarization contribution to the
    relaxation force (SPF, see main text). (c)~Collision force.
    (d,e)~Schematic depiction of the effect of image charges on the relaxation
    force, and the resulting SPF component parallel to the surface for
    dielectric mismatch $\Delta=-1$ (d) and $\Delta=1$~(e).}
  \label{fig:forces}
\end{figure}

An important advantage offered by particle-based modeling is that it permits
examination of the individual contributions to the forces exerted on ions near
the interface. Figure~\ref{fig:forces}a presents the total (i.e., arising from
ionic as well as induced charges) Coulombic force on ions as a function of
distance to the channel wall. As predicted, for attractive polarization the
magnitude of the relaxation force decreases near the wall, whereas for
repulsive polarization the magnitude of this force increases compared to the
case without dielectric mismatch.

Any asymmetry in the ionic atmosphere will be reflected in the surface
polarization charge. Thus, an interesting secondary effect arises, as this
surface polarization will also contribute to the relaxation force.  This
contribution, which we denote the \emph{surface polarization force} (SPF),
acts on ions near the wall and can be isolated in the simulations. Due to the
asymmetry of the ion cloud, the SPF has a nonzero component parallel to the
surface. Figure~\ref{fig:forces}b shows that for $\Delta = -1$ the SPF
diminishes the total relaxation force, whereas for $\Delta = 1$ it provides an
enhancement.  The reason for this is clarified by the schematics in
Fig.~\ref{fig:forces}d,e.  For $\Delta = -1$ (Fig.~\ref{fig:forces}d) the
image cloud carries a charge opposite to that of the ionic atmosphere, thus
causing a SPF in the direction of ion movement. For $\Delta = 1$
(Fig.~\ref{fig:forces}e) the ion cloud and its image carry the same charge, so
that the SPF opposes the ionic motion.  We observe that the SPF contribution
to the total relaxation force is considerably smaller for attractive surface
polarization than for the repulsive case, reflecting the weaker and less
asymmetric cloud in the first system.  Thus, the effect of surface
polarization on the relaxation force, and consequently on the ion mobility, is
twofold. First, it modifies the ion atmosphere and secondly, it exerts a
surface polarization force. Both of these effect diminish the relaxation force
when $\Delta = -1$ and enhance it when $\Delta = 1$.

Lastly, the distance dependence of the collision force opposing the ion
movement (Fig.~\ref{fig:forces}c) reflects the concentration profile,
increasing as more particles reside near the wall. However, as this force has
a weaker dependence on dielectric mismatch, the response of the
relaxation force dominates, giving rise to the counterintuitive behavior of
the mobility in Fig.~\ref{fig:ionatmosphere}a,b.

\begin{figure}
\includegraphics[scale=0.51]{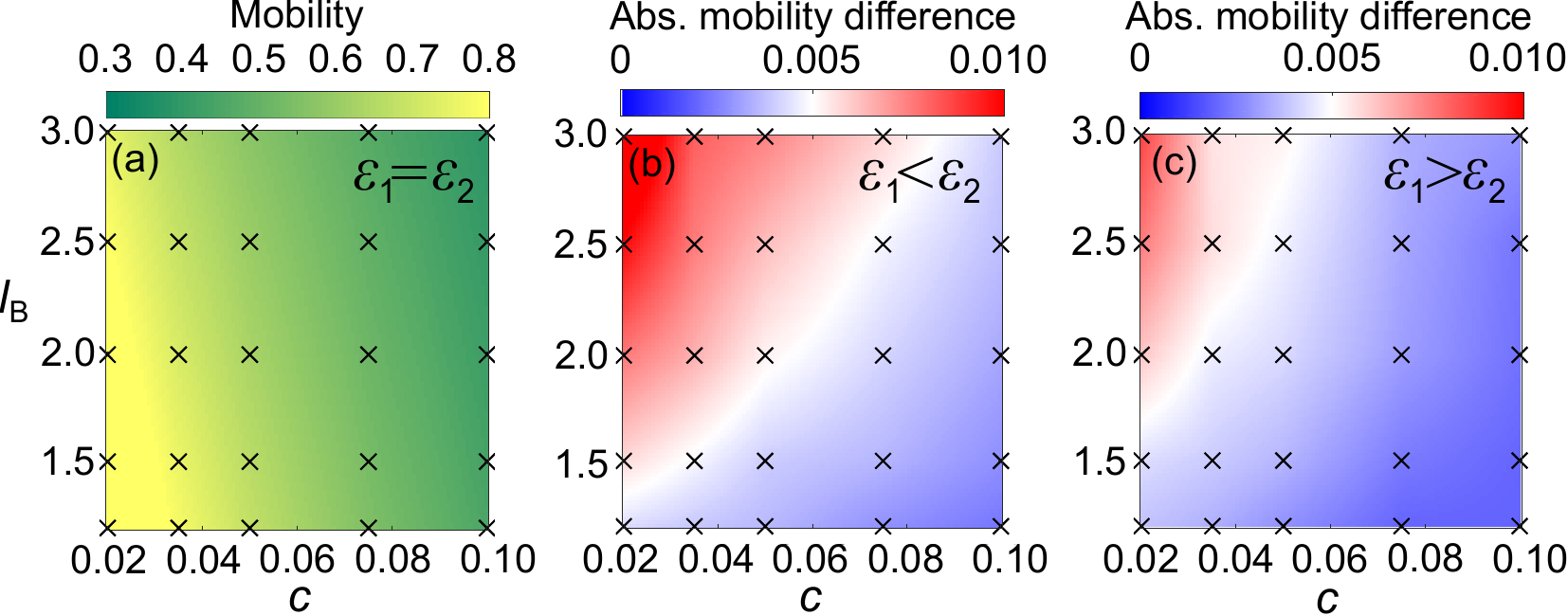}
\caption{Average ion mobilities as a function of Bjerrum length and ion
  concentration in the solution. The average is taken over all ions in the
  channel. Simulated values are marked with crosses; the color scheme
  results from 2D interpolation. a)~Mobility of ions in the absence of a
  dielectric mismatch, $\Delta = 0$. (b,c)~Absolute deviation in mobility
  compared to the $\Delta = 0$ situation for attractive polarization,
  $\Delta = -1$~(b) and repulsive polarization, $\Delta = 1$~(c). Units as in
  Fig.~\ref{fig:schematic}.}
\label{fig:2Dspeeds}
\end{figure}

The observations presented here depend on the global electrolyte concentration
and on the strength of the electrostatic coupling (expressed in terms of the
Bjerrum length $l_{\rm B}\propto(T\varepsilon_{1})^{-1}$), as those parameters
affect both bulk ion mobility and the screening of the surface
polarization. In Fig.~\ref{fig:2Dspeeds} we explore these dependencies. As a
baseline we employ the system without dielectric mismatch
(Fig.~\ref{fig:2Dspeeds}a), which confirms that the mobility decreases with
increasing concentration and increases with decreasing $l_{\rm B}$, as
expected~\cite{onsager31,fuoss57,fuoss78}.  Figures~\ref{fig:2Dspeeds}b,c show
the absolute deviations compared to this reference system for attractive and
repulsive surface polarization. We note that the effects of positive and
negative dielectric mismatch on ion mobility differ in magnitude. To emulate
an experimental set-up, the mobility in Fig.~\ref{fig:2Dspeeds} is determined
as an average across the entire channel. Thus, the suppressed electrolyte
concentration near low-permittivity surfaces (Fig.~\ref{fig:2Dspeeds}c)
diminishes the influence of reduced ion mobility on the observed average
mobility.

As the Bjerrum length is lowered, the region of significant mobility change is
reduced to lower concentrations. The lowest concentration studied here
is $0.02\sigma^{-3}$, corresponding to $0.1$M, i.e., comparable to
physiological salt concentrations. If concentrations are reduced further, the
effect of surface polarization is enhanced.

Our simulations lack a description of hydrodynamics beyond the Langevin
thermostat, and the long-range electrophoretic force is therefore absent in
our simulations~\cite{jardat99}. However, this force is affected by changes in
the ionic atmosphere in the same manner as the relaxation force, since the
magnitude of both forces is directly related to the amount and distribution of
charge within the ion cloud~\cite{onsager31}. Thus, inclusion of this force
should only enhance the phenomena reported here.
The use of an implicit solvent prevents us from observing effects related to
the molecular nature of the solvent. The hydration characteristics of ions can
affect their mobility by modulating the ion distribution near an
interface~\cite{qiao05a}. We also do not capture the effects of confinement on
the solvent structure, such as the formation of oriented hydration layers at
the channel edges and consequent slow-down of ions~\cite{qiao05a} due to
hindered water motion in these layers.  Moreover, such a layer would modify
the dielectric jump at the interface~\cite{bonthuis11}. Yet, the presence of a
hydration layer should not qualitatively affect the observed differences
between attractive and repulsive surface polarization.

Ion mobility and conductance in nanodevices are a delicate balance of several
contributions~\cite{balme15}, which along with the magnitude of the effect and
the nanometer scale of the devices may complicate experimental verification of
the dielectric modulation of ion mobilities.  This, however, does not mean
that this effect is of limited practical importance: it is amplified at low
concentration, permittivity, and temperature, and by high surface-to-volume
ratio.

In conclusion, we have demonstrated that the mobility of ions near interfaces
can be regulated via the dielectric mismatch between the solution and the wall
material. Surface polarization affects the mobility through two mechanisms,
both working in the same direction, that increase the mobility near a
high-permittivity surface and decrease it near a surface with low dielectric
constant.  First, surface polarization affects ion--ion interactions and
consequently the shape and intensity of the ionic atmosphere responsible for
the relaxation force. Secondly, due to the asymmetry of the counterion
atmosphere, a surface polarization force parallel to the interface emerges. We
anticipate that these findings can be exploited to understand and control
ionic flux on the nanoscale.

\section{Acknowledgements}

\begin{acknowledgments}
  We thank Jiaxing Yuan for the PPPM implementation of image charges. This
  work was supported by the National Institutes of Health through Grant No.\
  1R01 EB018358-01A1.  We acknowledge computational resources from the Quest
  high-performance computing facility at Northwestern University.
\end{acknowledgments}


\begin{thebibliography}{37}%
\makeatletter
\providecommand \@ifxundefined [1]{%
 \@ifx{#1\undefined}
}%
\providecommand \@ifnum [1]{%
 \ifnum #1\expandafter \@firstoftwo
 \else \expandafter \@secondoftwo
 \fi
}%
\providecommand \@ifx [1]{%
 \ifx #1\expandafter \@firstoftwo
 \else \expandafter \@secondoftwo
 \fi
}%
\providecommand \natexlab [1]{#1}%
\providecommand \enquote  [1]{``#1''}%
\providecommand \bibnamefont  [1]{#1}%
\providecommand \bibfnamefont [1]{#1}%
\providecommand \citenamefont [1]{#1}%
\providecommand \href@noop [0]{\@secondoftwo}%
\providecommand \href [0]{\begingroup \@sanitize@url \@href}%
\providecommand \@href[1]{\@@startlink{#1}\@@href}%
\providecommand \@@href[1]{\endgroup#1\@@endlink}%
\providecommand \@sanitize@url [0]{\catcode `\\12\catcode `\$12\catcode
  `\&12\catcode `\#12\catcode `\^12\catcode `\_12\catcode `\%12\relax}%
\providecommand \@@startlink[1]{}%
\providecommand \@@endlink[0]{}%
\providecommand \url  [0]{\begingroup\@sanitize@url \@url }%
\providecommand \@url [1]{\endgroup\@href {#1}{\urlprefix }}%
\providecommand \urlprefix  [0]{URL }%
\providecommand \Eprint [0]{\href }%
\providecommand \doibase [0]{http://dx.doi.org/}%
\providecommand \selectlanguage [0]{\@gobble}%
\providecommand \bibinfo  [0]{\@secondoftwo}%
\providecommand \bibfield  [0]{\@secondoftwo}%
\providecommand \translation [1]{[#1]}%
\providecommand \BibitemOpen [0]{}%
\providecommand \bibitemStop [0]{}%
\providecommand \bibitemNoStop [0]{.\EOS\space}%
\providecommand \EOS [0]{\spacefactor3000\relax}%
\providecommand \BibitemShut  [1]{\csname bibitem#1\endcsname}%
\let\auto@bib@innerbib\@empty
\bibitem [{\citenamefont {Gouaux}\ and\ \citenamefont
  {MacKinnon}(2005)}]{gouaux05}%
  \BibitemOpen
  \bibfield  {author} {\bibinfo {author} {\bibfnamefont {E.}~\bibnamefont
  {Gouaux}}\ and\ \bibinfo {author} {\bibfnamefont {R.}~\bibnamefont
  {MacKinnon}},\ }\bibfield  {title} {\enquote {\bibinfo {title} {Principles of
  selective ion transport in channels and pumps},}\ }\href {\doibase
  10.1126/science.1113666} {\bibfield  {journal} {\bibinfo  {journal}
  {Science}\ }\textbf {\bibinfo {volume} {310}},\ \bibinfo {pages} {1461--1465}
  (\bibinfo {year} {2005})}\BibitemShut {NoStop}%
\bibitem [{\citenamefont {Kreuer}(2014)}]{kreuer14}%
  \BibitemOpen
  \bibfield  {author} {\bibinfo {author} {\bibfnamefont {K.-D.}\ \bibnamefont
  {Kreuer}},\ }\bibfield  {title} {\enquote {\bibinfo {title} {Ion conducting
  membranes for fuel cells and other electrochemical devices},}\ }\href
  {\doibase 10.1021/cm402742u} {\bibfield  {journal} {\bibinfo  {journal}
  {Chem. Mater.}\ }\textbf {\bibinfo {volume} {26}},\ \bibinfo {pages}
  {361--380} (\bibinfo {year} {2014})}\BibitemShut {NoStop}%
\bibitem [{\citenamefont {Kohlrausch}(1879)}]{kohlrausch1879}%
  \BibitemOpen
  \bibfield  {author} {\bibinfo {author} {\bibfnamefont {F.}~\bibnamefont
  {Kohlrausch}},\ }\bibfield  {title} {\enquote {\bibinfo {title} {{Das
  electrische Leitungsverm{\"o}gen der w{\"a}sserigen L{\"o}sungen von den
  Hydraten und Salzen der leichten Metalle, sowie von Kupfervitriol,
  Zinkvitriol und Silbersalpeter}},}\ }\href {\doibase
  10.1002/andp.18782420102} {\bibfield  {journal} {\bibinfo  {journal} {Ann.
  Phys. (Leipzig)}\ }\textbf {\bibinfo {volume} {242}},\ \bibinfo {pages}
  {1--210} (\bibinfo {year} {1879})}\BibitemShut {NoStop}%
\bibitem [{\citenamefont {Kohlrausch}(1907)}]{kohlrausch07}%
  \BibitemOpen
  \bibfield  {author} {\bibinfo {author} {\bibfnamefont {F.}~\bibnamefont
  {Kohlrausch}},\ }\bibfield  {title} {\enquote {\bibinfo {title} {{\"U}ber
  {I}onenbeweglichkeiten im {W}asser},}\ }\href {\doibase
  10.1002/bbpc.19070132502} {\bibfield  {journal} {\bibinfo  {journal} {Z.
  Elektrochem.}\ }\textbf {\bibinfo {volume} {13}},\ \bibinfo {pages}
  {333--344} (\bibinfo {year} {1907})}\BibitemShut {NoStop}%
\bibitem [{\citenamefont {Atkins}\ and\ \citenamefont
  {de~Paula}(2006)}]{atkins2006}%
  \BibitemOpen
  \bibfield  {author} {\bibinfo {author} {\bibfnamefont {P.}~\bibnamefont
  {Atkins}}\ and\ \bibinfo {author} {\bibfnamefont {J.}~\bibnamefont
  {de~Paula}},\ }\href@noop {} {\emph {\bibinfo {title} {Physical
  Chemistry}}},\ \bibinfo {edition} {7th}\ ed.\ (\bibinfo  {publisher} {Oxford
  University Press},\ \bibinfo {address} {Oxford, U.K.},\ \bibinfo {year}
  {2006})\BibitemShut {NoStop}%
\bibitem [{\citenamefont {Debye}\ and\ \citenamefont
  {H{\"u}ckel}(1923)}]{debye23b}%
  \BibitemOpen
  \bibfield  {author} {\bibinfo {author} {\bibfnamefont {P.}~\bibnamefont
  {Debye}}\ and\ \bibinfo {author} {\bibfnamefont {E.}~\bibnamefont
  {H{\"u}ckel}},\ }\bibfield  {title} {\enquote {\bibinfo {title} {{Zur Theorie
  der Elektrolyte. II. Das Grenzgesetz f{\"u}r die elektrische
  Leitf{\"a}higkeit}},}\ }\href@noop {} {\bibfield  {journal} {\bibinfo
  {journal} {Phys. Z.}\ }\textbf {\bibinfo {volume} {24}},\ \bibinfo {pages}
  {305--325} (\bibinfo {year} {1923})}\BibitemShut {NoStop}%
\bibitem [{\citenamefont {Debye}(1927)}]{debye27}%
  \BibitemOpen
  \bibfield  {author} {\bibinfo {author} {\bibfnamefont {P.}~\bibnamefont
  {Debye}},\ }\bibfield  {title} {\enquote {\bibinfo {title} {Report on
  conductivity of strong electrolytes in dilute solutions},}\ }\href {\doibase
  10.1039/TF9272300334} {\bibfield  {journal} {\bibinfo  {journal} {Trans.
  Faraday Soc.}\ }\textbf {\bibinfo {volume} {23}},\ \bibinfo {pages}
  {334--340} (\bibinfo {year} {1927})}\BibitemShut {NoStop}%
\bibitem [{\citenamefont {Onsager}(1927)}]{onsager27}%
  \BibitemOpen
  \bibfield  {author} {\bibinfo {author} {\bibfnamefont {L.}~\bibnamefont
  {Onsager}},\ }\bibfield  {title} {\enquote {\bibinfo {title} {{Zur Theorie
  der Electrolyte. II}},}\ }\href@noop {} {\bibfield  {journal} {\bibinfo
  {journal} {Phys. Z.}\ }\textbf {\bibinfo {volume} {28}},\ \bibinfo {pages}
  {277--298} (\bibinfo {year} {1927})}\BibitemShut {NoStop}%
\bibitem [{\citenamefont {Onsager}\ and\ \citenamefont
  {Fuoss}(1931)}]{onsager31}%
  \BibitemOpen
  \bibfield  {author} {\bibinfo {author} {\bibfnamefont {L.}~\bibnamefont
  {Onsager}}\ and\ \bibinfo {author} {\bibfnamefont {R.~M.}\ \bibnamefont
  {Fuoss}},\ }\bibfield  {title} {\enquote {\bibinfo {title} {Irreversible
  processes in electrolytes. {D}iffusion, conductance, and viscous flow in
  arbitrary mixtures of strong electrolytes},}\ }\href {\doibase
  10.1021/j150341a001} {\bibfield  {journal} {\bibinfo  {journal} {J. Phys.
  Chem.}\ }\textbf {\bibinfo {volume} {36}},\ \bibinfo {pages} {2689--2778}
  (\bibinfo {year} {1931})}\BibitemShut {NoStop}%
\bibitem [{\citenamefont {Fuoss}\ and\ \citenamefont
  {Onsager}(1957)}]{fuoss57}%
  \BibitemOpen
  \bibfield  {author} {\bibinfo {author} {\bibfnamefont {R.~M.}\ \bibnamefont
  {Fuoss}}\ and\ \bibinfo {author} {\bibfnamefont {L.}~\bibnamefont
  {Onsager}},\ }\bibfield  {title} {\enquote {\bibinfo {title} {Conductance of
  unassociated electrolytes},}\ }\href@noop {} {\bibfield  {journal} {\bibinfo
  {journal} {J. Phys. Chem.}\ }\textbf {\bibinfo {volume} {61}},\ \bibinfo
  {pages} {668--682} (\bibinfo {year} {1957})}\BibitemShut {NoStop}%
\bibitem [{\citenamefont {Fuoss}(1978)}]{fuoss78}%
  \BibitemOpen
  \bibfield  {author} {\bibinfo {author} {\bibfnamefont {R.~M.}\ \bibnamefont
  {Fuoss}},\ }\bibfield  {title} {\enquote {\bibinfo {title}
  {Conductance-concentration function for the paired ion model},}\ }\href
  {\doibase 10.1021/j100511a017} {\bibfield  {journal} {\bibinfo  {journal} {J.
  Phys. Chem.}\ }\textbf {\bibinfo {volume} {82}},\ \bibinfo {pages}
  {2427--2440} (\bibinfo {year} {1978})}\BibitemShut {NoStop}%
\bibitem [{\citenamefont {Duan}\ and\ \citenamefont {Majumdar}(2010)}]{duan10}%
  \BibitemOpen
  \bibfield  {author} {\bibinfo {author} {\bibfnamefont {C.}~\bibnamefont
  {Duan}}\ and\ \bibinfo {author} {\bibfnamefont {A.}~\bibnamefont
  {Majumdar}},\ }\bibfield  {title} {\enquote {\bibinfo {title} {Anomalous ion
  transport in 2-nm hydrophilic nanochannels},}\ }\href@noop {} {\bibfield
  {journal} {\bibinfo  {journal} {Nature Nanotech.}\ }\textbf {\bibinfo
  {volume} {5}},\ \bibinfo {pages} {848--852} (\bibinfo {year}
  {2010})}\BibitemShut {NoStop}%
\bibitem [{\citenamefont {Stein}\ \emph {et~al.}(2004)\citenamefont {Stein},
  \citenamefont {Kruithof},\ and\ \citenamefont {Dekker}}]{stein04}%
  \BibitemOpen
  \bibfield  {author} {\bibinfo {author} {\bibfnamefont {D.}~\bibnamefont
  {Stein}}, \bibinfo {author} {\bibfnamefont {M.}~\bibnamefont {Kruithof}},\
  and\ \bibinfo {author} {\bibfnamefont {C.}~\bibnamefont {Dekker}},\
  }\bibfield  {title} {\enquote {\bibinfo {title} {Surface-charge-governed ion
  transport in nanofluidic channels},}\ }\href@noop {} {\bibfield  {journal}
  {\bibinfo  {journal} {Phys. Rev. Lett.}\ }\textbf {\bibinfo {volume} {93}},\
  \bibinfo {pages} {035901} (\bibinfo {year} {2004})}\BibitemShut {NoStop}%
\bibitem [{\citenamefont {Nishizawa}\ \emph {et~al.}(1995)\citenamefont
  {Nishizawa}, \citenamefont {Menon},\ and\ \citenamefont
  {Martin}}]{nishizawa95}%
  \BibitemOpen
  \bibfield  {author} {\bibinfo {author} {\bibfnamefont {M.}~\bibnamefont
  {Nishizawa}}, \bibinfo {author} {\bibfnamefont {V.~P.}\ \bibnamefont
  {Menon}},\ and\ \bibinfo {author} {\bibfnamefont {C.~R.}\ \bibnamefont
  {Martin}},\ }\bibfield  {title} {\enquote {\bibinfo {title} {Metal nanotubule
  membranes with electrochemically switchable ion-transport selectivity},}\
  }\href {\doibase 10.1126/science.268.5211.700} {\bibfield  {journal}
  {\bibinfo  {journal} {Science}\ }\textbf {\bibinfo {volume} {268}},\ \bibinfo
  {pages} {700--702} (\bibinfo {year} {1995})}\BibitemShut {NoStop}%
\bibitem [{\citenamefont {Cervera}\ \emph {et~al.}(2006)\citenamefont
  {Cervera}, \citenamefont {Schiedt}, \citenamefont {Neumann}, \citenamefont
  {Maf{\'e}},\ and\ \citenamefont {Ram{\'\i}rez}}]{cervera06}%
  \BibitemOpen
  \bibfield  {author} {\bibinfo {author} {\bibfnamefont {J.}~\bibnamefont
  {Cervera}}, \bibinfo {author} {\bibfnamefont {B.}~\bibnamefont {Schiedt}},
  \bibinfo {author} {\bibfnamefont {R.}~\bibnamefont {Neumann}}, \bibinfo
  {author} {\bibfnamefont {S.}~\bibnamefont {Maf{\'e}}},\ and\ \bibinfo
  {author} {\bibfnamefont {P.}~\bibnamefont {Ram{\'\i}rez}},\ }\bibfield
  {title} {\enquote {\bibinfo {title} {Ionic conduction, rectification, and
  selectivity in single conical nanopores},}\ }\href {\doibase
  10.1063/1.2179797} {\bibfield  {journal} {\bibinfo  {journal} {J. Chem.
  Phys.}\ }\textbf {\bibinfo {volume} {124}},\ \bibinfo {pages} {104706}
  (\bibinfo {year} {2006})}\BibitemShut {NoStop}%
\bibitem [{\citenamefont {Qiao}\ and\ \citenamefont
  {Aluru}(2005{\natexlab{a}})}]{qiao05b}%
  \BibitemOpen
  \bibfield  {author} {\bibinfo {author} {\bibfnamefont {R.}~\bibnamefont
  {Qiao}}\ and\ \bibinfo {author} {\bibfnamefont {N.~R.}\ \bibnamefont
  {Aluru}},\ }\bibfield  {title} {\enquote {\bibinfo {title} {Scaling of
  electrokinetic transport in nanometer channels},}\ }\href {\doibase
  10.1021/la0511900} {\bibfield  {journal} {\bibinfo  {journal} {Langmuir}\
  }\textbf {\bibinfo {volume} {21}},\ \bibinfo {pages} {8972--8977} (\bibinfo
  {year} {2005}{\natexlab{a}})}\BibitemShut {NoStop}%
\bibitem [{\citenamefont {Qiao}\ and\ \citenamefont
  {Aluru}(2005{\natexlab{b}})}]{qiao05a}%
  \BibitemOpen
  \bibfield  {author} {\bibinfo {author} {\bibfnamefont {R.}~\bibnamefont
  {Qiao}}\ and\ \bibinfo {author} {\bibfnamefont {N.~R.}\ \bibnamefont
  {Aluru}},\ }\bibfield  {title} {\enquote {\bibinfo {title} {Atomistic
  simulation of {KCl} transport in charged silicon nanochannels: {I}nterfacial
  effects},}\ }\href@noop {} {\bibfield  {journal} {\bibinfo  {journal}
  {Colloids Surf. A}\ }\textbf {\bibinfo {volume} {267}},\ \bibinfo {pages}
  {103--109} (\bibinfo {year} {2005}{\natexlab{b}})}\BibitemShut {NoStop}%
\bibitem [{\citenamefont {Buyukdagli}\ \emph {et~al.}(2011)\citenamefont
  {Buyukdagli}, \citenamefont {Manghi},\ and\ \citenamefont
  {Palmeri}}]{buyukdagli11}%
  \BibitemOpen
  \bibfield  {author} {\bibinfo {author} {\bibfnamefont {S.}~\bibnamefont
  {Buyukdagli}}, \bibinfo {author} {\bibfnamefont {M.}~\bibnamefont {Manghi}},
  \ and\ \bibinfo {author} {\bibfnamefont {J.}~\bibnamefont {Palmeri}},\
  }\bibfield  {title} {\enquote {\bibinfo {title} {Ionic exclusion phase
  transition in neutral and weakly charged cylindrical nanopores},}\
  }\href@noop {} {\bibfield  {journal} {\bibinfo  {journal} {J. Chem. Phys.}\
  }\textbf {\bibinfo {volume} {134}},\ \bibinfo {pages} {074706} (\bibinfo
  {year} {2011})}\BibitemShut {NoStop}%
\bibitem [{\citenamefont {Boda}\ \emph {et~al.}(2006)\citenamefont {Boda},
  \citenamefont {Valisk\'{o}}, \citenamefont {Eisenberg}, \citenamefont
  {Nonner}, \citenamefont {Henderson},\ and\ \citenamefont
  {Gillespie}}]{boda06}%
  \BibitemOpen
  \bibfield  {author} {\bibinfo {author} {\bibfnamefont {D.}~\bibnamefont
  {Boda}}, \bibinfo {author} {\bibfnamefont {M.}~\bibnamefont {Valisk\'{o}}},
  \bibinfo {author} {\bibfnamefont {B.}~\bibnamefont {Eisenberg}}, \bibinfo
  {author} {\bibfnamefont {W.}~\bibnamefont {Nonner}}, \bibinfo {author}
  {\bibfnamefont {D.}~\bibnamefont {Henderson}},\ and\ \bibinfo {author}
  {\bibfnamefont {D.}~\bibnamefont {Gillespie}},\ }\bibfield  {title} {\enquote
  {\bibinfo {title} {The effect of protein dielectric coefficient on the
  selectivity of a calcium channel},}\ }\href@noop {} {\bibfield  {journal}
  {\bibinfo  {journal} {J. Chem. Phys.}\ }\textbf {\bibinfo {volume} {125}},\
  \bibinfo {pages} {034901} (\bibinfo {year} {2006})}\BibitemShut {NoStop}%
\bibitem [{\citenamefont {Zhang}\ \emph {et~al.}(2011)\citenamefont {Zhang},
  \citenamefont {Ai}, \citenamefont {Liu}, \citenamefont {Joo},\ and\
  \citenamefont {Qian}}]{zhang11}%
  \BibitemOpen
  \bibfield  {author} {\bibinfo {author} {\bibfnamefont {B.}~\bibnamefont
  {Zhang}}, \bibinfo {author} {\bibfnamefont {Y.}~\bibnamefont {Ai}}, \bibinfo
  {author} {\bibfnamefont {J.}~\bibnamefont {Liu}}, \bibinfo {author}
  {\bibfnamefont {S.~W.}\ \bibnamefont {Joo}},\ and\ \bibinfo {author}
  {\bibfnamefont {S.}~\bibnamefont {Qian}},\ }\bibfield  {title} {\enquote
  {\bibinfo {title} {Polarization effect of a dielectric membrane on the ionic
  current rectification in a conical nanopore},}\ }\href@noop {} {\bibfield
  {journal} {\bibinfo  {journal} {J. Phys. Chem. C}\ }\textbf {\bibinfo
  {volume} {115}},\ \bibinfo {pages} {24951--24959} (\bibinfo {year}
  {2011})}\BibitemShut {NoStop}%
\bibitem [{\citenamefont {Mamonov}\ \emph {et~al.}(2003)\citenamefont
  {Mamonov}, \citenamefont {Coalson}, \citenamefont {Nitzan},\ and\
  \citenamefont {Kurnikova}}]{mamonov03}%
  \BibitemOpen
  \bibfield  {author} {\bibinfo {author} {\bibfnamefont {A.~B.}\ \bibnamefont
  {Mamonov}}, \bibinfo {author} {\bibfnamefont {R.~D.}\ \bibnamefont
  {Coalson}}, \bibinfo {author} {\bibfnamefont {A.}~\bibnamefont {Nitzan}},\
  and\ \bibinfo {author} {\bibfnamefont {M.~G.}\ \bibnamefont {Kurnikova}},\
  }\bibfield  {title} {\enquote {\bibinfo {title} {The role of the dielectric
  barrier in narrow biological channels: {A} novel composite approach to
  modeling single-channel currents},}\ }\href {\doibase
  http://doi.org/10.1016/S0006-3495(03)75095-4} {\bibfield  {journal} {\bibinfo
   {journal} {Biophys. J.}\ }\textbf {\bibinfo {volume} {84}},\ \bibinfo
  {pages} {3646--3661} (\bibinfo {year} {2003})}\BibitemShut {NoStop}%
\bibitem [{\citenamefont {Tagliazucchi}\ \emph {et~al.}(2013)\citenamefont
  {Tagliazucchi}, \citenamefont {Rabin},\ and\ \citenamefont
  {Szleifer}}]{tagliazucchi13}%
  \BibitemOpen
  \bibfield  {author} {\bibinfo {author} {\bibfnamefont {M.}~\bibnamefont
  {Tagliazucchi}}, \bibinfo {author} {\bibfnamefont {Y.}~\bibnamefont {Rabin}},
  \ and\ \bibinfo {author} {\bibfnamefont {I.}~\bibnamefont {Szleifer}},\
  }\bibfield  {title} {\enquote {\bibinfo {title} {Transport rectification in
  nanopores with outer membranes modified with surface charges and
  polyelectrolytes},}\ }\href {\doibase 10.1021/nn403686s} {\bibfield
  {journal} {\bibinfo  {journal} {ACS Nano}\ }\textbf {\bibinfo {volume} {7}},\
  \bibinfo {pages} {9085--9097} (\bibinfo {year} {2013})}\BibitemShut {NoStop}%
\bibitem [{\citenamefont {Balme}\ \emph {et~al.}(2015)\citenamefont {Balme},
  \citenamefont {Picaud}, \citenamefont {Manghi}, \citenamefont {Palmeri},
  \citenamefont {Bechelany}, \citenamefont {Cabello-Aguilar}, \citenamefont
  {Abou-Chaaya}, \citenamefont {Miele}, \citenamefont {Balanzat},\ and\
  \citenamefont {Janot}}]{balme15}%
  \BibitemOpen
  \bibfield  {author} {\bibinfo {author} {\bibfnamefont {S.}~\bibnamefont
  {Balme}}, \bibinfo {author} {\bibfnamefont {F.}~\bibnamefont {Picaud}},
  \bibinfo {author} {\bibfnamefont {M.}~\bibnamefont {Manghi}}, \bibinfo
  {author} {\bibfnamefont {J.}~\bibnamefont {Palmeri}}, \bibinfo {author}
  {\bibfnamefont {M.}~\bibnamefont {Bechelany}}, \bibinfo {author}
  {\bibfnamefont {S.}~\bibnamefont {Cabello-Aguilar}}, \bibinfo {author}
  {\bibfnamefont {A.}~\bibnamefont {Abou-Chaaya}}, \bibinfo {author}
  {\bibfnamefont {P.}~\bibnamefont {Miele}}, \bibinfo {author} {\bibfnamefont
  {E.}~\bibnamefont {Balanzat}},\ and\ \bibinfo {author} {\bibfnamefont
  {J.~M.}\ \bibnamefont {Janot}},\ }\bibfield  {title} {\enquote {\bibinfo
  {title} {Ionic transport through sub-10 nm diameter hydrophobic high-aspect
  ratio nanopores: experiment, theory and simulation},}\ }\href@noop {}
  {\bibfield  {journal} {\bibinfo  {journal} {Sci. Rep.}\ }\textbf {\bibinfo
  {volume} {5}},\ \bibinfo {pages} {10135} (\bibinfo {year}
  {2015})}\BibitemShut {NoStop}%
\bibitem [{\citenamefont {Luijten}\ \emph {et~al.}(2002)\citenamefont
  {Luijten}, \citenamefont {Fisher},\ and\ \citenamefont
  {Panagiotopoulos}}]{luijten02a}%
  \BibitemOpen
  \bibfield  {author} {\bibinfo {author} {\bibfnamefont {E.}~\bibnamefont
  {Luijten}}, \bibinfo {author} {\bibfnamefont {M.~E.}\ \bibnamefont {Fisher}},
  \ and\ \bibinfo {author} {\bibfnamefont {A.~Z.}\ \bibnamefont
  {Panagiotopoulos}},\ }\bibfield  {title} {\enquote {\bibinfo {title}
  {Universality class of criticality in the restricted primitive model
  electrolyte},}\ }\href@noop {} {\bibfield  {journal} {\bibinfo  {journal}
  {Phys. Rev. Lett.}\ }\textbf {\bibinfo {volume} {88}},\ \bibinfo {pages}
  {185701} (\bibinfo {year} {2002})}\BibitemShut {NoStop}%
\bibitem [{\citenamefont {Neumann}(1883)}]{neumann1883}%
  \BibitemOpen
  \bibfield  {author} {\bibinfo {author} {\bibfnamefont {C.}~\bibnamefont
  {Neumann}},\ }\href@noop {} {\emph {\bibinfo {title} {Hydrodynamische
  {U}ntersuchungen, nebst einem {A}nhange {\"u}ber die {P}robleme der
  {E}lektrostatik und der magnetischen {I}nduction}}}\ (\bibinfo  {publisher}
  {B.G. Teubner},\ \bibinfo {address} {Leipzig},\ \bibinfo {year}
  {1883})\BibitemShut {NoStop}%
\bibitem [{\citenamefont {Hsiao}\ and\ \citenamefont
  {Luijten}(2006)}]{hsiao06}%
  \BibitemOpen
  \bibfield  {author} {\bibinfo {author} {\bibfnamefont {P.-Y.}\ \bibnamefont
  {Hsiao}}\ and\ \bibinfo {author} {\bibfnamefont {E.}~\bibnamefont
  {Luijten}},\ }\bibfield  {title} {\enquote {\bibinfo {title} {Salt-induced
  collapse and reexpansion of highly charged flexible polyelectrolytes},}\
  }\href@noop {} {\bibfield  {journal} {\bibinfo  {journal} {Phys. Rev. Lett.}\
  }\textbf {\bibinfo {volume} {97}},\ \bibinfo {pages} {148301} (\bibinfo
  {year} {2006})}\BibitemShut {NoStop}%
\bibitem [{not()}]{note-units}%
  \BibitemOpen
  \href@noop {} {}\bibinfo {note} {Using $\sigma=0.7$~nm one obtains
  $E\sim10^{7}$~V/m, whereas equating the ion mass to the mass of hydrated
  sodium (131.4 g/mol) yields $\tau\sim5.6$~ps. The damping constant in our
  simulations is about 2 orders of magnitude smaller than would be
  representative for the viscosity of water. This allows a faster exploration
  of the system configurations.}\BibitemShut {Stop}%
\bibitem [{\citenamefont {Wien}(1928)}]{wien28}%
  \BibitemOpen
  \bibfield  {author} {\bibinfo {author} {\bibfnamefont {M.}~\bibnamefont
  {Wien}},\ }\bibfield  {title} {\enquote {\bibinfo {title} {{\"U}ber den
  {S}pannungseffekt der {L}eitf{\"a}higkeit von {E}lektrolyten in niedrigeren
  {F}eldern},}\ }\href {\doibase 10.1002/andp.19283900704} {\bibfield
  {journal} {\bibinfo  {journal} {Ann. Phys. (Leipzig)}\ }\textbf {\bibinfo
  {volume} {390}},\ \bibinfo {pages} {795--811} (\bibinfo {year}
  {1928})}\BibitemShut {NoStop}%
\bibitem [{\citenamefont {Onsager}\ and\ \citenamefont
  {Kim}(1957)}]{onsager57}%
  \BibitemOpen
  \bibfield  {author} {\bibinfo {author} {\bibfnamefont {L.}~\bibnamefont
  {Onsager}}\ and\ \bibinfo {author} {\bibfnamefont {S.~K.}\ \bibnamefont
  {Kim}},\ }\bibfield  {title} {\enquote {\bibinfo {title} {Wien effect in
  simple strong electrolytes},}\ }\href {\doibase 10.1021/j150548a015}
  {\bibfield  {journal} {\bibinfo  {journal} {J. Phys. Chem.}\ }\textbf
  {\bibinfo {volume} {61}},\ \bibinfo {pages} {198--215} (\bibinfo {year}
  {1957})}\BibitemShut {NoStop}%
\bibitem [{\citenamefont {Luijten}(2013)}]{luijten13}%
  \BibitemOpen
  \bibfield  {author} {\bibinfo {author} {\bibfnamefont {E.}~\bibnamefont
  {Luijten}},\ }\bibfield  {title} {\enquote {\bibinfo {title}
  {Electrochemistry: Discrete answer},}\ }\href@noop {} {\bibfield  {journal}
  {\bibinfo  {journal} {Nature Phys.}\ }\textbf {\bibinfo {volume} {9}},\
  \bibinfo {pages} {606--607} (\bibinfo {year} {2013})}\BibitemShut {NoStop}%
\bibitem [{\citenamefont {Debye}\ and\ \citenamefont
  {Falkenhagen}(1928)}]{debyefalk28}%
  \BibitemOpen
  \bibfield  {author} {\bibinfo {author} {\bibfnamefont {P.}~\bibnamefont
  {Debye}}\ and\ \bibinfo {author} {\bibfnamefont {H.}~\bibnamefont
  {Falkenhagen}},\ }\bibfield  {title} {\enquote {\bibinfo {title} {{Dispersion
  der Leitf{\"a}higkeit und der Dielektrizit{\"a}tskonstante starker
  Elektrolyte}},}\ }\href@noop {} {\bibfield  {journal} {\bibinfo  {journal}
  {Phys. Z.}\ }\textbf {\bibinfo {volume} {29}},\ \bibinfo {pages} {401--426}
  (\bibinfo {year} {1928})}\BibitemShut {NoStop}%
\bibitem [{\citenamefont {Falkenhagen}(1934)}]{falkenhagen1934}%
  \BibitemOpen
  \bibfield  {author} {\bibinfo {author} {\bibfnamefont {H.}~\bibnamefont
  {Falkenhagen}},\ }\href {https://books.google.com/books?id=FZozAAAAIAAJ}
  {\emph {\bibinfo {title} {Electrolytes}}}\ (\bibinfo  {publisher} {Clarendon
  Press},\ \bibinfo {address} {Oxford},\ \bibinfo {year} {1934})\BibitemShut
  {NoStop}%
\bibitem [{\citenamefont {Nadler}\ \emph {et~al.}(2003)\citenamefont {Nadler},
  \citenamefont {Hollerbach},\ and\ \citenamefont {Eisenberg}}]{nadler03}%
  \BibitemOpen
  \bibfield  {author} {\bibinfo {author} {\bibfnamefont {B.}~\bibnamefont
  {Nadler}}, \bibinfo {author} {\bibfnamefont {U.}~\bibnamefont {Hollerbach}},
  \ and\ \bibinfo {author} {\bibfnamefont {R.~S.}\ \bibnamefont {Eisenberg}},\
  }\bibfield  {title} {\enquote {\bibinfo {title} {Dielectric boundary force
  and its crucial role in gramicidin},}\ }\href {\doibase
  10.1103/PhysRevE.68.021905} {\bibfield  {journal} {\bibinfo  {journal} {Phys.
  Rev. E}\ }\textbf {\bibinfo {volume} {68}},\ \bibinfo {pages} {021905}
  (\bibinfo {year} {2003})}\BibitemShut {NoStop}%
\bibitem [{\citenamefont {Zwanikken}\ and\ \citenamefont {Olvera de~la
  Cruz}(2013)}]{zwanikken13}%
  \BibitemOpen
  \bibfield  {author} {\bibinfo {author} {\bibfnamefont {J.~W.}\ \bibnamefont
  {Zwanikken}}\ and\ \bibinfo {author} {\bibfnamefont {M.}~\bibnamefont {Olvera
  de~la Cruz}},\ }\bibfield  {title} {\enquote {\bibinfo {title} {Tunable soft
  structure in charged fluids confined by dielectric interfaces},}\ }\href@noop
  {} {\bibfield  {journal} {\bibinfo  {journal} {Proc. Natl. Acad. Sci.
  U.S.A.}\ }\textbf {\bibinfo {volume} {110}},\ \bibinfo {pages} {5301--5308}
  (\bibinfo {year} {2013})}\BibitemShut {NoStop}%
\bibitem [{\citenamefont {Korobeynikov}\ \emph {et~al.}(2005)\citenamefont
  {Korobeynikov}, \citenamefont {Melekhov}, \citenamefont {Soloveitchik},
  \citenamefont {Royak}, \citenamefont {Agoris},\ and\ \citenamefont
  {Pyrgioti}}]{korobeynikov05}%
  \BibitemOpen
  \bibfield  {author} {\bibinfo {author} {\bibfnamefont {S.~M.}\ \bibnamefont
  {Korobeynikov}}, \bibinfo {author} {\bibfnamefont {A.~V.}\ \bibnamefont
  {Melekhov}}, \bibinfo {author} {\bibfnamefont {Y.~G.}\ \bibnamefont
  {Soloveitchik}}, \bibinfo {author} {\bibfnamefont {M.~E.}\ \bibnamefont
  {Royak}}, \bibinfo {author} {\bibfnamefont {D.~P.}\ \bibnamefont {Agoris}},\
  and\ \bibinfo {author} {\bibfnamefont {E.}~\bibnamefont {Pyrgioti}},\
  }\bibfield  {title} {\enquote {\bibinfo {title} {Surface conductivity at the
  interface between ceramics and transformer oil},}\ }\href
  {http://stacks.iop.org/0022-3727/38/i=6/a=021} {\bibfield  {journal}
  {\bibinfo  {journal} {J. Phys. D: Appl. Phys.}\ }\textbf {\bibinfo {volume}
  {38}},\ \bibinfo {pages} {915--921} (\bibinfo {year} {2005})}\BibitemShut
  {NoStop}%
\bibitem [{\citenamefont {Jardat}\ \emph {et~al.}(1999)\citenamefont {Jardat},
  \citenamefont {Bernard}, \citenamefont {Turq},\ and\ \citenamefont
  {Kneller}}]{jardat99}%
  \BibitemOpen
  \bibfield  {author} {\bibinfo {author} {\bibfnamefont {M.}~\bibnamefont
  {Jardat}}, \bibinfo {author} {\bibfnamefont {O.}~\bibnamefont {Bernard}},
  \bibinfo {author} {\bibfnamefont {P.}~\bibnamefont {Turq}},\ and\ \bibinfo
  {author} {\bibfnamefont {G.~R.}\ \bibnamefont {Kneller}},\ }\bibfield
  {title} {\enquote {\bibinfo {title} {Transport coefficients of electrolyte
  solutions from {S}mart {B}rownian dynamics simulations},}\ }\href {\doibase
  10.1063/1.478703} {\bibfield  {journal} {\bibinfo  {journal} {J. Chem.
  Phys.}\ }\textbf {\bibinfo {volume} {110}},\ \bibinfo {pages} {7993--7999}
  (\bibinfo {year} {1999})}\BibitemShut {NoStop}%
\bibitem [{\citenamefont {Bonthuis}\ \emph {et~al.}(2011)\citenamefont
  {Bonthuis}, \citenamefont {Gekle},\ and\ \citenamefont {Netz}}]{bonthuis11}%
  \BibitemOpen
  \bibfield  {author} {\bibinfo {author} {\bibfnamefont {D.~J.}\ \bibnamefont
  {Bonthuis}}, \bibinfo {author} {\bibfnamefont {S.}~\bibnamefont {Gekle}},\
  and\ \bibinfo {author} {\bibfnamefont {R.~R.}\ \bibnamefont {Netz}},\
  }\bibfield  {title} {\enquote {\bibinfo {title} {Dielectric profile of
  interfacial water and its effect on double-layer capacitance},}\ }\href
  {\doibase 10.1103/PhysRevLett.107.166102} {\bibfield  {journal} {\bibinfo
  {journal} {Phys. Rev. Lett.}\ }\textbf {\bibinfo {volume} {107}},\ \bibinfo
  {pages} {166102} (\bibinfo {year} {2011})}\BibitemShut {NoStop}%
\end{thebibliography}

%

\end{document}